\documentclass[12pt,a4paper,dvips]{article}
\usepackage{a4p}
\usepackage{cite,mcite}
\usepackage{graphicx}
\usepackage{physics}
\usepackage{l3_title}
\usepackage{ifthen}
\usepackage{Lep}
\journalname{Phys. Lett. B}
\date{November 2, 1997}
\preprint{99-156}
\Lep{1}
\newlength{\capindent}
\newlength{\capwidth}
\newlength{\figwidth}
\newcommand{\icaption}[2][!*!,!]{\hspace*{\capindent}%
  \begin{minipage}{\capwidth}
    \ifthenelse{\equal{#1}{!*!,!}}%
      {\caption{#2}}%
      {\caption[#1]{#2}}
  \end{minipage}}
\newcommand {\ngbcc}  {\overline{n}_{\rm{g} \rarrow \ccbar}}
\newcommand {\ngbbb}  {\overline{n}_{\rm{g} \rarrow \bbbar}}
\newcommand {\ngbqq}  {\overline{n}_{\rm{g} \rarrow \rm{Q}\overline{\rm{Q}}}}

\begin{document}
\begin{titlepage}
\date{\today}

\title{Measurement of the Probability of Gluon Splitting into
 Charmed Quarks in Hadronic
 Z Decays}
\author{The L3 Collaboration}
%
%
\begin{abstract}

We have measured the probability, $\ngbcc$,
 of a gluon splitting into a charm-quark pair 
using 1.7 million hadronic Z decays collected by the L3 
detector. Two independent methods have been applied to events with a
 three-jet topology. 
One method relies on tagging charmed hadrons by identifying a lepton in the 
lowest energy jet. 
The other method uses a neural network based on global
event shape parameters.  
Combining both methods, we measure   
$\ngbcc = [2.45 \pm 0.29 \pm 0.53]\%$.

\end{abstract}
%
%
\submitted

\end{titlepage}

\section{Introduction}
\begin{figure} 
 \begin{center}
   \includegraphics[width=12cm]{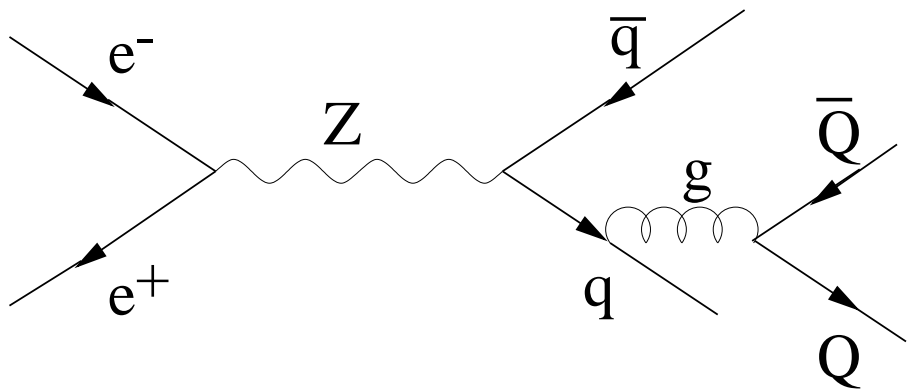}
 \end{center}
 \caption[fig1]{The lowest-order Feynman diagram for a gluon splitting into
 heavy-quark pairs, where Q represents a charm or bottom quark.}
 \label{fyen}
\end{figure}

The process of a gluon splitting into heavy quark pairs in hadronic Z 
decays, as shown in Fig. \ref{fyen}, provides a good test 
of perturbative QCD~\cite{mana,sey1,sey2,misey}. 
The energy scale of heavy quark pair production
is 
in the perturbative QCD region, and the process is infrared safe because
of the natural cutoff provided by the heavy quark mass. It should be
mentioned that only the process of open heavy quark production
is considered here. Measurements of hidden charm ($\rm{J}/\psi$) production
via QCD processes in Z decays have also been made ~\cite{hiddc}.

The average number of heavy-quark pairs produced by gluon  
splitting per hadronic Z decay is defined by
\begin{eqnarray} 
 \ngbqq = \frac{N(\rm{Z} \ra \qqbar\rm{g},~~\rm{g} \ra
 \rm{Q}\overline{\rm{Q}})}
                         {N(\rm{Z} \ra \rm{hadrons})}~~, 
\end{eqnarray}
where Q stands for a charm or bottom quark. 
The most recent theoretical predictions\cite{misey} to leading order 
in $\alpha_{\rm{s}}$, obtained by resumming large leading and
 next-to-leading logarithmic terms to 
all orders, give $\ngbcc \approx 2\%$ and 
$\ngbbb \approx 0.2\%$.  
There are previous experimental results for 
$\ngbcc$~\cite{opal1,opal2,opal3,aleph1} and
$\ngbbb$~\cite{del1,alp1,del2} . 
The uncertainty of these measurements contributes to 
the systematic error of electroweak measurements related to heavy quarks 
and is the biggest single source of systematic error in 
the measurement of $R_{\rm{b}} = \Gamma_{\bbbar} /
 \Gamma_{\rm{had}}$\cite{rb}. 
Therefore, a good knowledge of $\overline{n}_{\rm{g} \ra \ccbar}$ and  
$\overline{n}_{\rm{g} \ra \bbbar}$ is important for precision tests of the 
Standard Model in the heavy-quark sector. 
In this letter, we present a measurement of $\overline{n}_{\rm{g} \ra \ccbar}$
using a high statistics sample of hadronic Z decay events collected 
by L3~\cite{x10}  during the years 1994 and 1995. 
To identify events with a gluon jet, we first select three-jet events  
by applying a jet-finding algorithm that optimises the yield of 
$\rm{g} \ra \ccbar$ events.     
Subsequently, two methods are used to identify charm quarks from gluon
 splitting. 
The first method, the lepton analysis, searches for a lepton 
from semi-leptonic charm decays in the lowest 
energy jet. According to the JETSET Parton Shower~\cite{x11} event generator,
 the latter has 
about an $80\%$  probability of being a gluon jet for the identified three jet
sample.
The second method, the hadronic event shape analysis, uses 
a neural network technique, 
with input nodes consisting of several global event shape 
variables, to distinguish  $ \rm{g} \ra \rm{Q} \overline{\rm{Q}} $ events 
from backgrounds. 

\section{Three-Jet Event Selection}
 Hadronic events are selected by criteria similiar to the ones used for the
measurement of the total hadronic cross section\cite{x12}. 
The number of selected hadronic events is 1.74 million, with an
estimated  background of 
$0.15\%$ from other processes. 
To identify the gluon jet, we require a three-jet event 
topology and assume that the lowest energy jet is a gluon jet.
The jets are found by the JADE algorithm~\cite{x13}
with a $y_{cut}$ value of 0.03, which maximises the fraction of 
$\rm{g} \ra \ccbar$ events. This cut assigns 
 63\% of $\rm{g} \ra \ccbar$ events
to the three-jet topology. 
The jet energies are calculated using the 
relation:

\begin{eqnarray} 
  E_{i}  =  E_{\rm{cm}}\left[ \frac{ \sin \psi_{jk}}
          {\sin \psi_{jk}+\sin \psi_{ki}+\sin \psi_{ij}}\right]~~,
\end{eqnarray} 
where $E_{\rm{cm}}$ is the center of mass energy and $\psi_{ij}$ is the angle in space 
between jets $i$ and $j$. Planar (three jet) events are defined by the condition that 
 the sum of the angles between the three jets is  greater than 
$358^{\circ}$. The jets are labelled in decreasing order of jet energy.
To ensure that most charged particles are contained in the acceptance 
 of the Silicon Microvertex Detector which is used to improve the
 event signature, the  polar
angle of the thrust axis, $\theta_{\rm{T}}$, must satisfy the condition:
  $|\cos \theta_{\rm{T}}| < 0.7 $.
With  these additional criteria, 430k hadronic events are selected.

\section{Lepton Analysis}

\subsection{$\rm{g} \ra \ccbar$ Event Selection and Background Sources}
This method uses events with an electron or a muon candidate 
in the lowest energy jet to tag a charm quark from gluon 
splitting. Electron candidates are selected by the criteria:
\newline 
\begin{itemize}
\item An energy cluster with electromagnetic energy greater than 
 2 GeV, hadronic energy less than 2 GeV and polar angle, $\theta$, such that
 $|\cos \theta|< 0.72 $.
\item The electromagnetic energy cluster must have between 10 and
 40 associated BGO crystals and more than 95$\%$ of the energy in the
 9 central crystals.
\item Matching is required in the azimuthal angle $\phi$ and transverse
 energy between the energy cluster and a track in the central track chamber.
\end{itemize}
For the muon candidates it is required that:
\begin{itemize}
\item A track is found in the barrel muon system with hits in 2 or
 more $\phi$ chamber layers and 1 or more Z-chamber layers.
\item The distance of closest approach to the fill vertex in the
 transverse plane is less than 100 mm, as well as less than three
 times the estimated  error due to multiple scattering in the
 calorimeters. 
\end{itemize} 
 More details of the electron and muon selection criteria
are given in reference~\cite{x14}.
The electron energy is required to be between 2 GeV and 6 GeV, 
while the measured muon momentum must be between 3 GeV and 6 GeV;   
the lepton selection gives 1000 electron and 1287 muon candidates.
For background and acceptance studies, a Monte Carlo 
sample of 5.2 million hadronic Z decays is used. 
The events are generated using the JETSET 7.3 generator\cite{x11},
 passed through the full L3 detector simulation~\cite{x15} and analysed in the same 
way as the data.
The following backgrounds, to the $\rm{g} \ra \ccbar$ signal contribute, 
in decreasing order of 
importance, to 
the selected lepton samples: 

\begin{itemize}
    \item[-] $\rm{Z} \ra \bbbar$ and $\rm{Z} \ra \ccbar$ events with hard gluon
 radiation.
 In this case, the lowest energy jet contains a lepton coming from a
 semi-leptonic decay of a primary 
heavy quark.
    \item[-] Hadrons misidentified as lepton candidates.
    \item[-] Decay in flight of $\pi^{\pm}$ or $\rm{K}^{\pm}$ into a muon.
    \item[-] Dalitz decays: $\pi^{0},\eta \ra \ee \gamma$.
    \item[-] Photon conversions into electron pairs.
   \item[-] $\rm{Z} \ra \qqbar \rm{g},~ \rm{g} \ra \bbbar$.
\end{itemize}
  
 Monte Carlo studies show that $62 \%$ of the selected events with an electron and
$47 \%$ of the events with a muon contain a directly produced heavy quark, 
usually a b or $\bbar$, in the lowest energy jet. 
To reduce this background, we use an event discriminant 
variable\cite{x16}, $\it D$, the negative logarithm of the probability that
all tracks originate from the primary vertex, which
has small values when all tracks
originate from the primary vertex and large values for events
containing secondary vertices.
Applying the cut $D < 1.5$
 rejects $58\%$ of the b-quark background while keeping 
$85\%$ of the $\rm{g} \ra \ccbar$ events. 
Further discrimination between gluon splitting events and directly produced
b or $\bbar$ quarks is provided by the invariant mass of the lowest 
energy jet, $M_{j3}$;
events are rejected if $M_{j3} <~6.5~\rm{GeV}$. 

\begin{figure}
 \begin{center}
   \includegraphics*[width=8cm,]{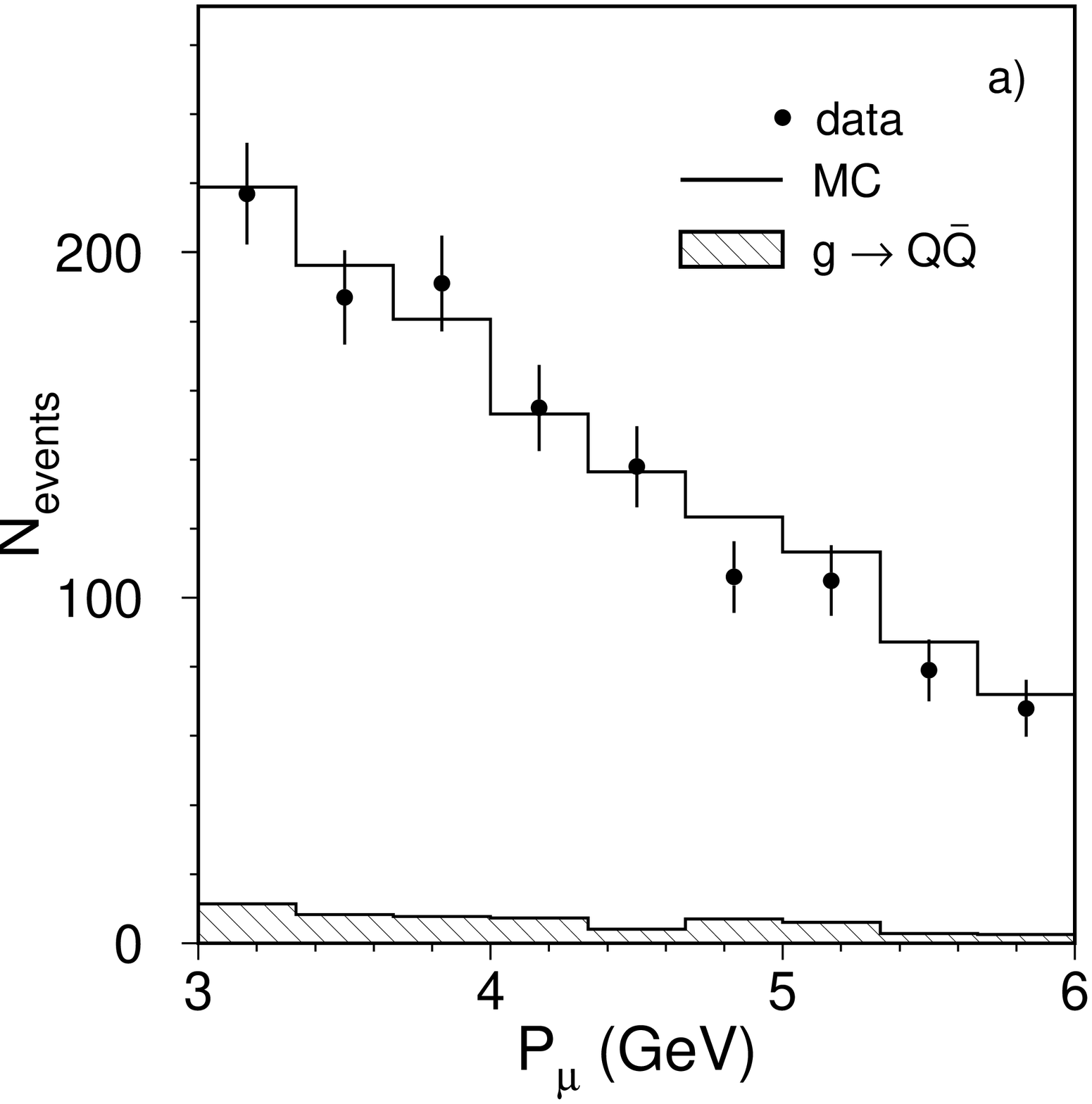} 
   \includegraphics*[width=8cm]{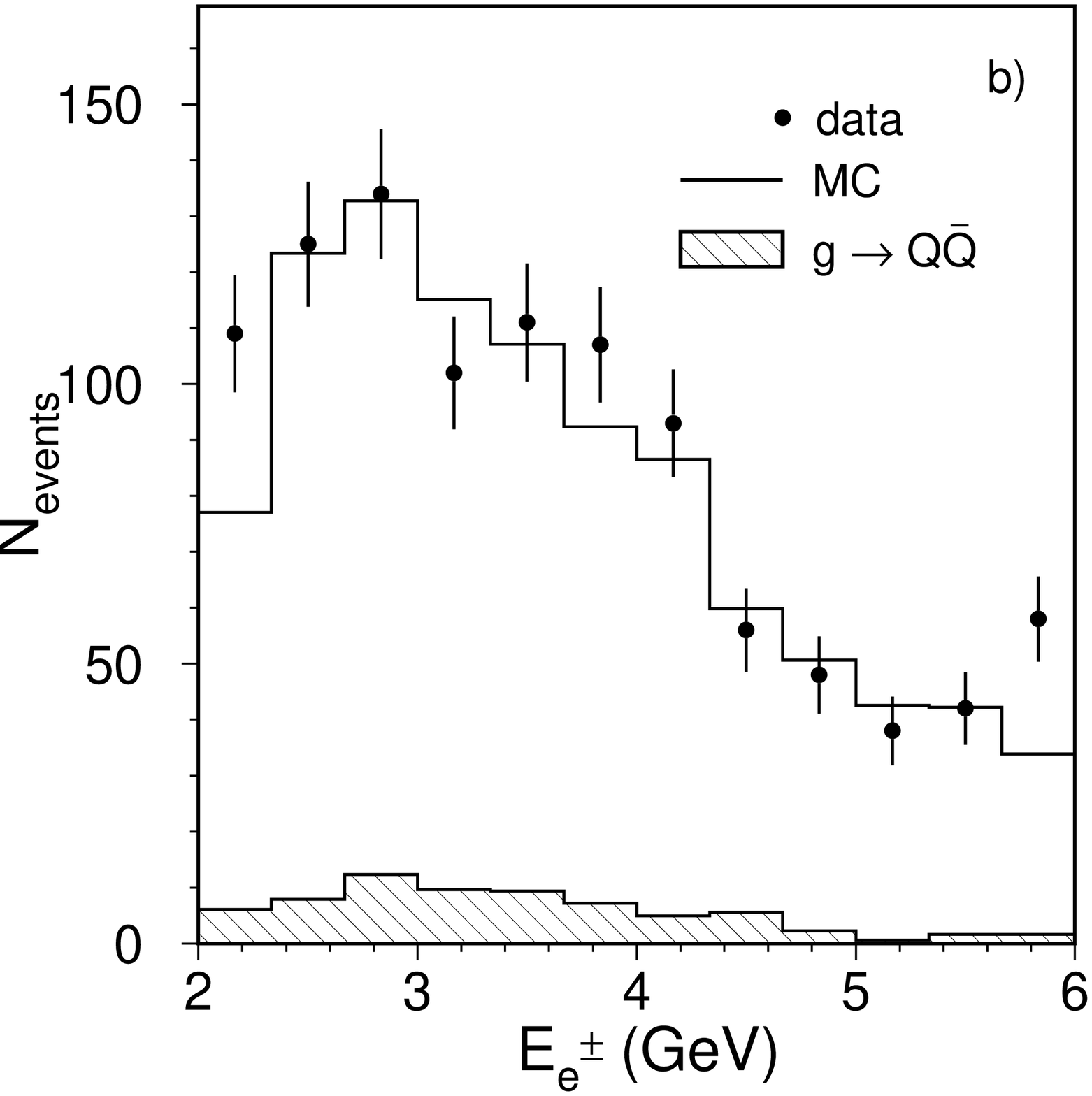}\\
\vspace{1cm}   
   \includegraphics*[width=8cm]{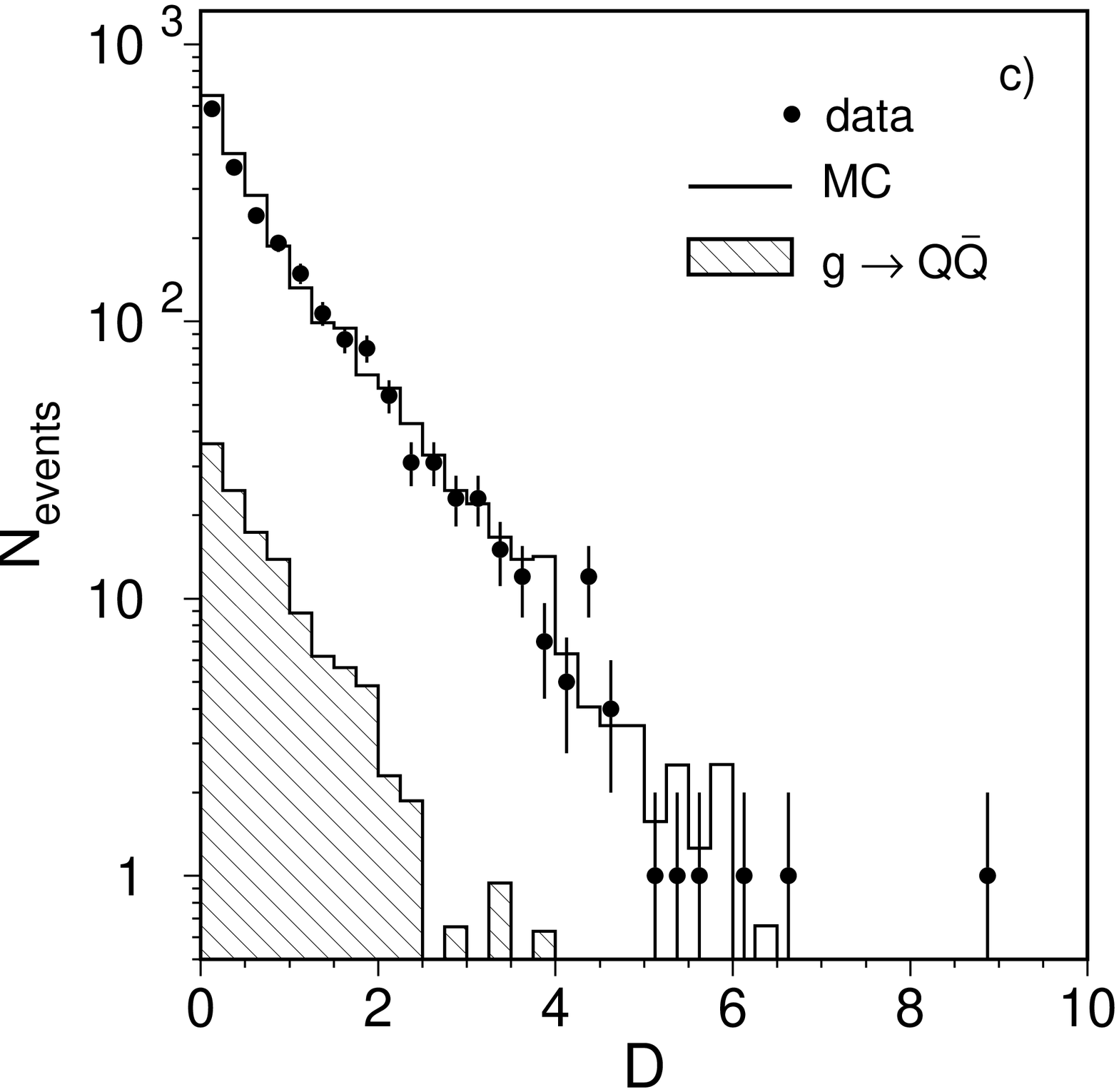}
   \includegraphics*[width=8cm]{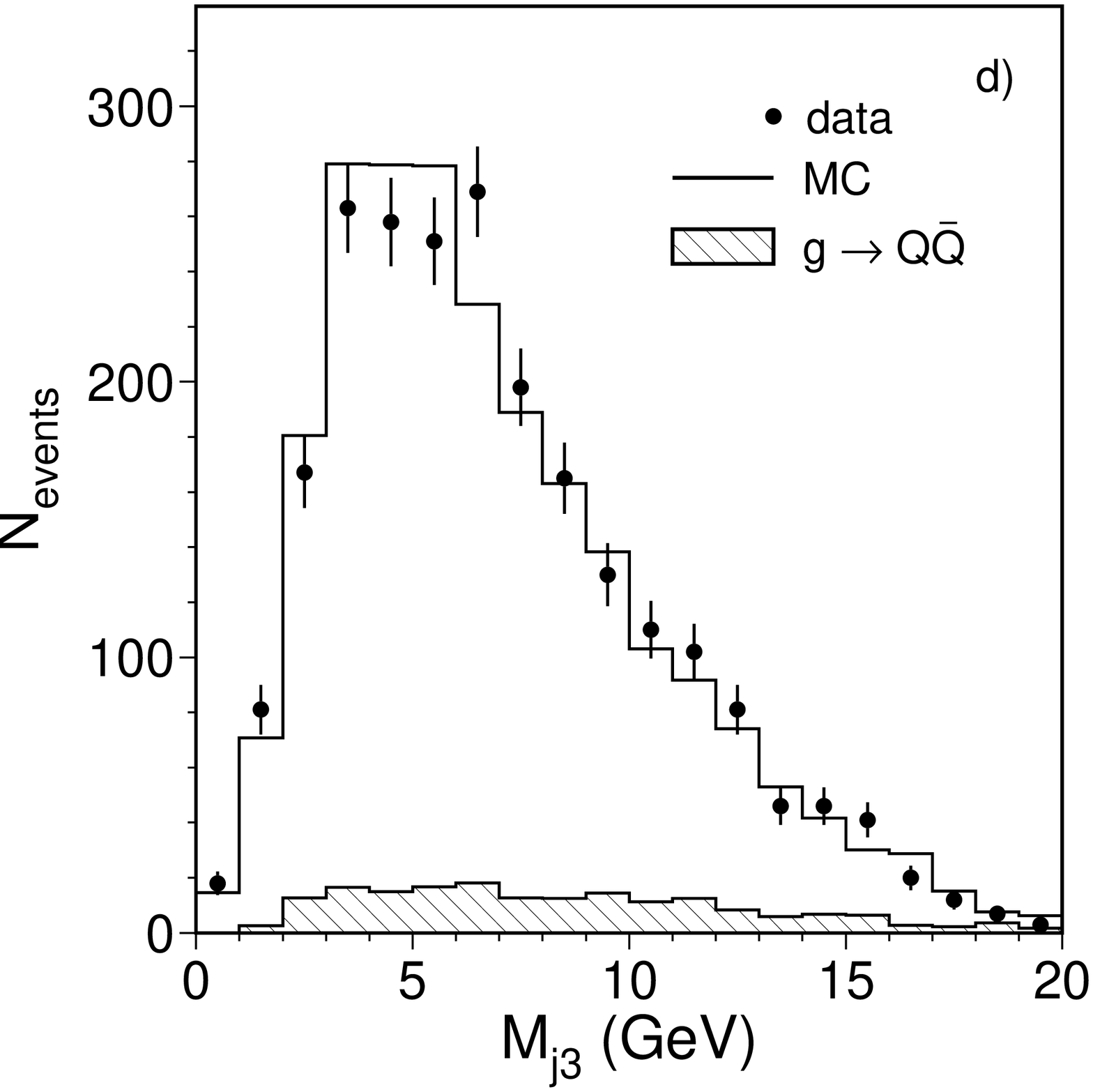}
 \end{center}
\caption{Distributions of: a) the momentum of 
the muon candidates and b) the energy of electron candidates in the lowest energy jet, 
c) the event discriminant, $\it D$, and
d) the invariant mass of the lowest energy jet. Data are points with error bars and
the open histogram is the JETSET Monte Carlo prediction. 
 The shaded histogram shows the 
 JETSET prediction for $\rm{g} \rightarrow \rm{Q} \overline{\rm{Q}}$.  
}
\label{mcck}
\end{figure}

Since all backgrounds are estimated from Monte Carlo events, we have 
compared several of the crucial Monte Carlo distributions 
with the data. For example, the muon and electron momentum, the event discriminant $\it D$, and
 the effective mass of the lowest energy jet, are shown in Fig.~\ref{mcck}a-d. Good
agreement is seen for all these distributions.  

In order to estimate the background from jet misassignment and  
to check the heavy flavour composition in the data and Monte Carlo, we extract 
the fraction of $\Zo \ra \bbbar$ events ($R_{\rm{b}}^{3jet}$)
 in the three-jet events 
from data samples using a double-hemisphere tagging method\cite{x16}.
 The result found is
 $R_{\rm{b}}^{3jet} = 0.2017 \pm 0.0027 \pm 0.0030$, where the first error is
statistical and the second is systematic, mainly due to uncertainties on  
the hemisphere correlations and on the selection efficiencies of 
the light and charm quarks. This is significantly different from the
untuned Monte Carlo
value of $0.2146 \pm 0.0003$. 
The $\Zo \ra \bbbar$ event fraction in the Monte Carlo is therefore corrected by
 reweighting the events according to the measured $R_{\rm{b}}^{3jet}$ value. 
The uncertaintity in the value of $R_{\rm{b}}^{3jet}$ is taken into account in 
the systematic error.

Applying the event discriminant and the third-jet 
invariant mass cut, the ratio between data and Monte Carlo predictions for 
the fraction of selected leptons in the two highest energy jets is found to be
consistent with unity:
\begin{eqnarray}        
 \frac{f^{DATA}_{\mu}}{f^{MC}_{\mu}} = 0.977 \pm 0.026,
 ~~\frac{f^{DATA}_{\rm{e}}}{f^{MC}_{\rm{e}}} = 1.025 \pm 0.053. \nonumber
\end{eqnarray} 
 In this background enriched sample good agreement is found between data and
 Monte Carlo. 
After all selection criteria, 360 electron
and 450 muon candidate events are selected in the lowest energy jet. 
The remaining backgrounds are estimated from Monte Carlo as shown in 
Table 1. 

\subsection{Results} 
After subtracting all background contributions 
except that due to $\rm{g} \rarrow \bbbar$, $51 \pm 24$ and 
$75 \pm 21$ $\rm{g} \ra \rm{Q}\overline{\rm{Q}}$ splitting events are found for 
the muon and electron channels, respectively, where the errors are statistical
only.
The average number of charm quark pairs from gluon splitting per hadronic event
is then given by:
\begin{eqnarray}  
 \ngbcc= \frac{ N_{sel} }
 {N_{had} \cdot \varepsilon^{\rm{c}} \cdot 2 \cdot Br(\rm{c} \rarrow \rm{X  \ell} \nu) }
 -\frac{\varepsilon^{\rm{b}} Br(\rm{b} \rarrow \rm{X \ell} \nu)}
{\varepsilon^{\rm{c}} Br(\rm{c} \rarrow \rm{X \ell} \nu)} \ngbbb, 
\end{eqnarray}
 Where $\varepsilon^{\rm{c}}$, $\varepsilon^{\rm{b}}$ are the selection efficiencies for~
 $\rm{g} \rarrow \ccbar$ and $\rm{g} \rarrow \bbbar$ events, respectively, given in Table 2,
 $N_{sel}$ is the number of events after background subtraction, 
$N_{had}$ is the total number of hadronic $ \Zo $ decays and
 $Br(\rm{c} \ra X  \ell \nu)$  and  $Br(\rm{b} \ra X  \ell \nu)$ are the c and b
 hadron semileptonic branching ratios~\cite{x18},
 given in Table 3.
Using the efficiencies shown in Table 2, we obtain:
\begin{eqnarray}   
\ngbcc &=& [3.06 \pm 1.76-2.59(
\ngbbb -0.26)]\%~~~(\rm{muons}) \nonumber \\
\ngbcc &=&[3.14 \pm 1.14-3.59(
\ngbbb -0.26)]\%~~~(\rm{electrons}) \nonumber 
\end{eqnarray}  
where the errors are from data statistics only. These results correspond to
the weighted average of the published values of $\ngbbb$~\cite{del1,alp1}
 of [$0.26 \pm 0.06$]$\%$.

\subsection{Systematic Errors}
 The different systematic errors for the lepton analysis are presented
 in Table 3. There are three distinct sources: 
 experimental systematic errors, errors from leptonic branching
 ratio uncertainties and
 modelling errors
 that can  affect both signal and background.
 The errors labelled `Monte Carlo Statistics' originate 
from the limited statistics of the Monte Carlo samples 
of the processes $\rm{g} \ra \ccbar$ and $\rm{g} \ra \bbbar$ used to calculate
 the selection efficiencies $\varepsilon^c$ and $\varepsilon^b$.
 For each cut or parameter variation all the efficiencies (including
 $\varepsilon^b$) were recalculated and used in Eqn.(3) to obtain the values
 of $\delta(\ngbcc)/\ngbcc$
 presented in Table 3.
 \par The errors on the lepton detection efficiency are estimated by varying
 the lepton selection cuts by $\pm 10\%$ around their nominal values.
 The errors on the background simulation are estimated using `lepton
 enriched' samples of electrons and muons. The systematic errors are
 estimated as the difference between data and
 Monte Carlo after applying the
 invariant mass and event discriminant cuts to these samples.
 This difference in the background due to misidentified hadrons,
 photon conversions, Dalitz decays and decays in flight amounts
 to 2.0$\%$ for muons and 8.1$\%$ for electrons.
 The `Track 
 Smearing' error is estimated by smearing tracks to improve
 the data/Monte Carlo comparison for the event discriminant. The error is the
 difference in the gluon splitting rate calculated with, or without, the 
 smeared Monte Carlo. The modelling errors are estimated by varying the
 parameters of the JETSET parton shower model that was previously tuned
 to our data~\cite{x19}. We vary, within
their measured errors, 
 the parameters $\epsilon_{b}$ and $\epsilon_{c}$
in the Peterson fragmentation function~\cite{x20}, the parameter $b$ in the Lund symmetric 
fragmentation function~\cite{x21} for light quarks, the parameter $\sigma_{q}$, describing
hadron transverse momenta and the QCD scale parameter $\Lambda_{LLA}$
used for the parton shower evolution. 
Since fully simulated events are not available with different values of
these parameters, their effects are determined from generated events 
with a detector resolution smeared so as to be consistent with the data. 
The systematic error due to different models of semileptonic decays of charm 
and bottom is also considered. 
In addition to the JETSET decay model, we estimate the uncertainty
by also considering the models of Altarelli \etal (ACCMM)\cite{x22}, 
Isgur \etal (ISGW) and the modified Isgur model (ISGW**)~\cite{x23}. 
The ACCMM model is used for the central values of our measurement.

\par Combining the muon and electron channels, taking into
account the correlated errors arising from the leptonic 
branching ratios and modelling,
 the overall systematic error is estimated 
to be $21.5 \% $, giving the result:   
\begin{eqnarray} 
\ngbcc &=& [3.12 \pm 0.96 \pm 0.67-3.28(
\ngbbb -0.26
)]\%~~(\rm{leptons}) \nonumber 
\end{eqnarray} 
where the first error is statistical and the
second systematic.

\section{Event Shape Analysis}
This analysis uses characteristic differences in event shape
distributions for the gluon splitting to heavy 
quark process, as compared to the background processes
where heavy quarks are directly produced by electroweak Z decays.
The rather low purity attainable as compared to the lepton analysis
is largely compensated by data statistics as all hadronic decays
of heavy quarks are utilised.

\subsection{$\rm{g} \ra \rm{Q} \overline{\rm{Q}}$ Event Selection}
In events with a three-jet topology, most of the gluon jets radiated by a primary 
quark are in the two lowest energy jets.        
Due to the large mass of the heavy quark, a gluon jet containing a heavy 
quark has a larger invariant mass and energy than one containing light quarks. 
Hence, the distribution of the sum of the invariant masses of the two lowest 
energy jets and the energy fraction in a cone around the jet axis are 
different in a gluon jet with heavy quarks than in one with light quarks. 
Observables, sensitive to the correlation among the three jet momenta, can 
also allow a discrimination between events with a gluon splitting to heavy 
quark pairs and to light quarks or to gluons. 
In this method, we use three different categories of Monte Carlo event samples
 as follows:
\newline
\begin{itemize}
 \item 80,000 events containing gluon splitting to charm quark pairs,  
   called the $\it C$ sample. 
 \item 8,000 events with gluon splitting to bottom quark pairs, called 
 the $\it B$ sample.
 \item 5.2 million events without the gluon splitting to 
heavy quark pair process, called the $\it N$ sample.
\end{itemize}
 
\begin{figure} 
 \begin{center}
   \includegraphics[width=14cm]{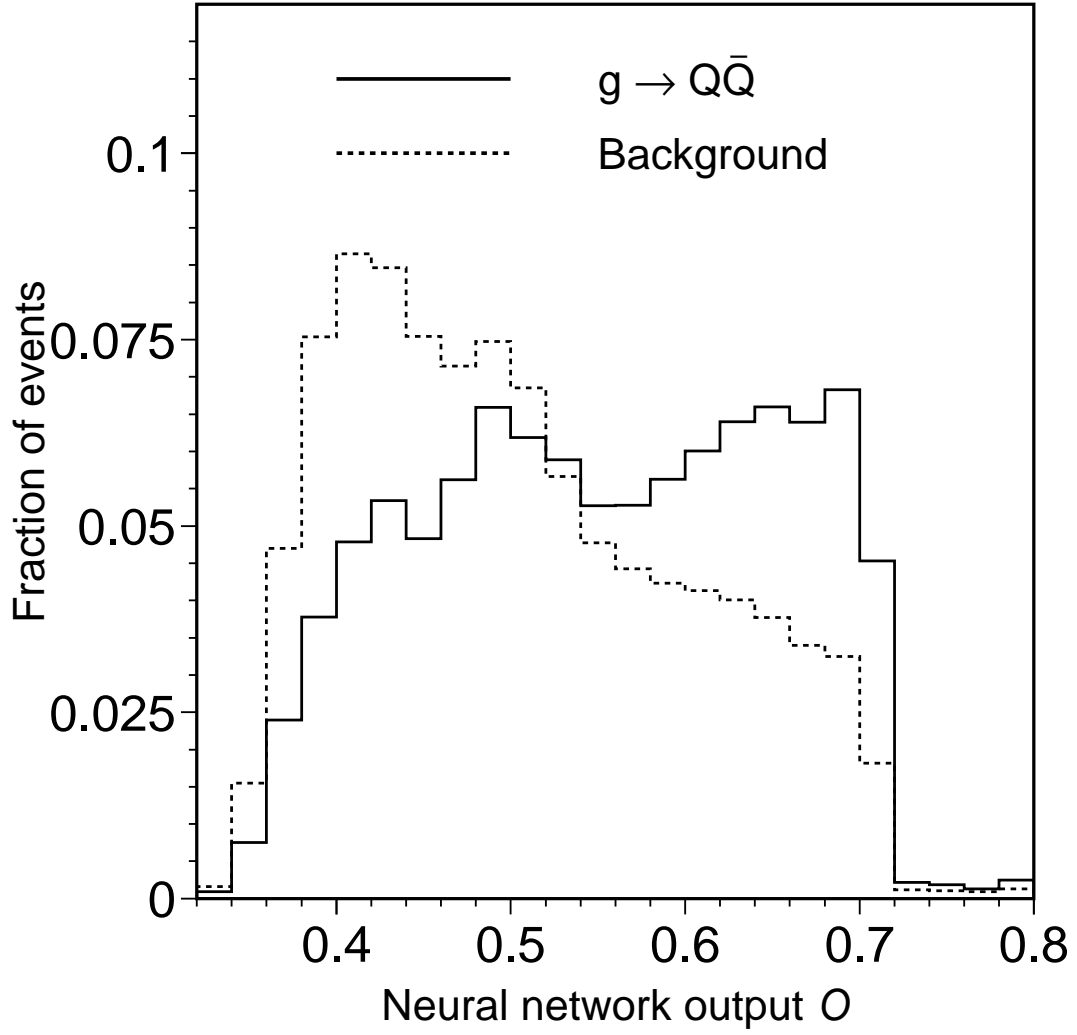}
 \end{center}
 \caption{ Neural network output distribution for $\rm{g} \ra \rm{Q}\overline{\rm{Q}}$ 
 (solid line) and
background events (dashed line). The distributions are normalised to the same area.
}
\label{nnef}
\end{figure}

 A neural network~\cite{jetnet} has been constructed based on the
 following five variables:

\begin{itemize}
\item The difference between the sum of the invariant masses of the two lowest
energy jets and the invariant mass of the highest energy jet,
 \(\Delta m \equiv m_{3} + m_{2} - m_{1} \nonumber\), 
where $m_{i}$ is the effective mass of jet $\it i$.
\item The energy in a cone of 8 degree half-angle 
around the jet axis of the second jet, divided by the
energy of the jet.
\item Three different
Fox-Wolfram moments ($\Pi_{1}$,$\Pi_{2}$ and $\Pi_{3}$),
calculated from the jet momenta, sensitive
to the global event topology, as described 
in \cite{x24}.
\end{itemize}
The neural network has five input nodes, one hidden layer of 10 nodes, and
one output node. 
For training, we assume that the samples $\it C$ and $\it B$ 
are the signal events and the $\it N$ sample is background. Subsamples of
10,000 $\it N$ events, 9,000 $\it C$ events and 1,000 $\it B$ events are used
for training.
Since events with a gluon splitting into light quark pairs or
 gluon pairs are
a major background source, we use the event discriminant 
variable, $\it D$, to reduce this background.   
Introducing a lower cut on the event discriminant variable in this
 analysis, the light quark background 
is reduced and the data sample is almost uncorrelated with the one used in 
the lepton analysis. 
Fig.\ref{nnef} shows the distribution of the neural network output, $O$,
for $\rm{g} \ra \rm{Q}\overline{\rm{Q}}$ and for background 
after applying the cut $\it D  >$ 1.0.  
Optimising the total error on the signal, we choose the 
 cut: $O > 0.59 $.
 The corresponding selection efficiencies of the data and of the three
 Monte Carlo 
samples are listed in Table~\ref{tab:nns}. 
   Requiring only the three-jet event selection, 
 the purity for the $\rm{g} \ra \ccbar$ events is found to be $2.5\%$.
After the cuts $O > 0.59$ and $\it{D} >$ 1.0, the purity is
increased to $4.5 \%$ and systematic 
uncertainties are reduced.
 The purity estimations given correspond to the value of 0.15$\%$
 for $\ngbbb$ used in the JETSET generator.
In the region, $O > 0.59$ and $1.0 <\it D <$ 1.5, only 
35 events from the lepton 
analysis are found among the 9,520 selected events. The lepton and 
event shape analyses can therefore be considered as uncorrelated. 

\subsection{Results and Systematic Errors} 
 We extract the 
probability of a gluon splitting into charm quark pairs using the  
relation:
\begin{equation}  
 \ngbcc = \frac{ \varepsilon_{D} - \varepsilon_{N} 
 -(\varepsilon_{B} - \varepsilon_{N}) \ngbbb}
{\varepsilon_{C} - \varepsilon_{N} }~~,
\end{equation}
where $\varepsilon_{D}$, $\varepsilon_{C}$, $\varepsilon_{B}$ and $\varepsilon_{N}$ are 
the efficiencies for the data, $\it C$, $\it B$ 
and $\it N$ samples, respectively.
 The values obtained are listed in Table~\ref{tab:nns} and they yield the 
 result:
\begin{eqnarray}
\ngbcc &=& [2.27 \pm 0.30-3.86(
\ngbbb -0.26)]\%, \nonumber
\end{eqnarray}  
where the error is from data statistics only, and the dependence on $\ngbbb$
is shown explicitly.
  This value is in good agreement with that 
obtained by the lepton analysis.
\par The statistical significance of the observed signal is illustrated in
 Fig 4.
 The distributions of the neural network output for the data 
and the background Monte Carlo are shown in Fig 4a. The relative 
normalisations are determined according to Eqn.(4) in the region
$O > 0.59$. The corresponding $\rm{g} \rightarrow\rm{Q} \overline{\rm{Q}}$ signal
after background subtraction is shown in Fig 4b as compared
to the Monte Carlo prediction normalised to the measured
value of $\ngbcc$. Although the analysis
has a low purity, the gluon splitting signal is seen to be large.
 For the region $O > 0.59$, it amounts to 2064 $\pm$ 182 events.
The hatched area in Fig 4b shows the estimated 
$\rm{g} \rarrow \bbbar$ contribution.
\par Systematic errors in the efficiencies arise due to imperfect modelling
of the event shape distributions in the Monte Carlo.
 We estimate the systematic errors due to several
input parameters in the JETSET Parton Shower generator by varying their values 
within the estimated errors. For this purpose, fast simulations which take into 
account the detector resolution have been performed for $\it N$, $\it C$ and 
$\it B$ samples.   
Comparing the efficiencies of the new samples, generated by variation of one of
 the parameters, to those obtained 
 with the tuned parameter from the reference sample, the changes  
$\delta(\ngbcc)/\ngbcc$, due to
the uncertainties of different parameters, are found and listed
in Table~\ref{tab:evsh}.    

In addition to the tuned fragmentation parameters in JETSET,
we consider the following sources of 
systematic error:
\newline
\begin{itemize}  
 \item Monte Carlo statistics.
  
\item   $R_{\rm{b}}^{3jet}$: We correct the Monte Carlo 
samples by the measured value of $R_{\rm{b}}^{3jet}$ in three-jet events. 
The changes due to the measurement error on this quantity are assigned 
as a systematic error.

\item Track smearing: This error is estimated in the same way as for the
 leptonic analysis (see section 3.3 above).

\item Energy calibration: since we reconstruct three-jet events with
calorimetric clusters, the systematic uncertainty from the energy calibration
of the calorimeters is studied. We use the Monte Carlo samples
of  reconstructed
three-jet events with energy smearing of clusters 
according to the nominal calorimeter resolution.
We repeat the analysis and assign the difference as the error. 
\end{itemize}
Our measurement is also checked by using different values of 
$y_{cut}$.  
With the values of 0.02 and 0.04 of $y_{cut}$, we measure the   
$\ngbcc$ to be $(2.48 \pm 0.39)\%$ and
$(2.03 \pm 0.27)\%$, respectively,
where the error is only statistical. These values are very consistent
with that given by the standard value $y_{cut} =$ 0.03. The stability
of the result to variation of the event discriminant and neural 
network cuts was also investigated. No systematic deviations were
observed.  
Adding all the systematic errors in quadrature, we obtain a total systematic
error of 23.7\%, yielding the result: 

\begin{eqnarray} 
\ngbcc &=& [2.27 \pm 0.30 \pm  0.54-3.86(
\ngbbb -0.26)]\%~~(\rm{event shape}). \nonumber
\end{eqnarray}

\section{Combined Result and Discussion}
The measurements from the two different methods are
now combined in an uncorrelated manner giving the result:
\begin{eqnarray} 
\ngbcc &=& [2.45 \pm 0.29 \pm 0.53 ]\%~~, \nonumber
\end{eqnarray} 
 where it is assumed that $\overline{n}_{\rm{g} \rarrow \bbbar} = [0.26 \pm 0.06]\%$. 
 In the errors quoted, statistical and systematic errors (assumed uncorrelated in
 the two different analysis methods) are combined in quadrature.  
 Our result is compared with other measurements and theoretical 
predictions in Table 6. Good agreement is found between the 
different experimental measurements. The latest resummed perturbative QCD
calculation~\cite{misey} agrees, within errors, with our measurement.
Our result is also in good agreement with the prediction of 
the ARIADNE and JETSET generators.

%
\section*{Acknowledgements}

    We express our gratitude to the CERN accelerator divisions for
    the excellent performance of the LEP machine.
    We acknowledge with appreciation the effort of all engineers,
    technicians and support staff who have participated in the
    construction and maintenance of this experiment.
%

%
\newpage
\typeout{   }     
\typeout{Using author list for paper 194 -?}
\typeout{$Modified: Wed Oct 27 09:16:14 1999 by clare $}
\typeout{!!!!  This should only be used with document option a4p!!!!}
\typeout{   }
%
%
%
%
%
%

\newcount\tutecount  \tutecount=0
\def\tutenum#1{\global\advance\tutecount by 1 \xdef#1{\the\tutecount}}
\def\tute#1{$^{#1}$}
\tutenum\aachen            
\tutenum\nikhef            
\tutenum\mich              
\tutenum\lapp              
\tutenum\basel             
\tutenum\lsu               
\tutenum\beijing           
\tutenum\berlin            
\tutenum\bologna           
\tutenum\tata              
\tutenum\ne                
\tutenum\bucharest         
\tutenum\budapest          
\tutenum\mit               
\tutenum\debrecen          
\tutenum\florence          
\tutenum\cern              
\tutenum\wl                
\tutenum\geneva            
\tutenum\hefei             
\tutenum\seft              
\tutenum\lausanne          
\tutenum\lecce             
\tutenum\lyon              
\tutenum\madrid            
\tutenum\milan             
\tutenum\moscow            
\tutenum\naples            
\tutenum\cyprus            
\tutenum\nymegen           
\tutenum\caltech           
\tutenum\perugia           
\tutenum\cmu               
\tutenum\prince            
\tutenum\rome              
\tutenum\peters            
\tutenum\salerno           
\tutenum\ucsd              
\tutenum\santiago          
\tutenum\sofia             
\tutenum\korea             
\tutenum\alabama           
\tutenum\utrecht           
\tutenum\purdue            
\tutenum\psinst            
\tutenum\zeuthen           
\tutenum\eth               
\tutenum\hamburg           
\tutenum\taiwan            
\tutenum\tsinghua          
{
\parskip=0pt
\noindent
{\bf The L3 Collaboration:}
\ifx\selectfont\undefined
 \baselineskip=10.8pt
 \baselineskip\baselinestretch\baselineskip
 \normalbaselineskip\baselineskip
 \ixpt
\else
 \fontsize{9}{10.8pt}\selectfont
\fi
\medskip
\tolerance=10000
\hbadness=5000
\raggedright
\hsize=162truemm\hoffset=0mm
\def\r{\rlap,}
\noindent

M.Acciarri\r\tute\milan\
P.Achard\r\tute\geneva\ 
O.Adriani\r\tute{\florence}\ 
M.Aguilar-Benitez\r\tute\madrid\ 
J.Alcaraz\r\tute\madrid\ 
G.Alemanni\r\tute\lausanne\
J.Allaby\r\tute\cern\
A.Aloisio\r\tute\naples\ 
M.G.Alviggi\r\tute\naples\
G.Ambrosi\r\tute\geneva\
% ...
\newpage
\begin{list}{A}{\itemsep=0pt plus 0pt minus 0pt\parsep=0pt plus 0pt minus 0pt
                \topsep=0pt plus 0pt minus 0pt}
\item[\aachen]
 I. Physikalisches Institut, RWTH, D-52056 Aachen, FRG$^{\S}$\\
 III. Physikalisches Institut, RWTH, D-52056 Aachen, FRG$^{\S}$
\item[\nikhef] National Institute for High Energy Physics, NIKHEF, 
     and University of Amsterdam, NL-1009 DB Amsterdam, The Netherlands
\item[\mich] University of Michigan, Ann Arbor, MI 48109, USA
\item[\lapp] Laboratoire d'Annecy-le-Vieux de Physique des Particules, 
     LAPP,IN2P3-CNRS, BP 110, F-74941 Annecy-le-Vieux CEDEX, France
\item[\basel] Institute of Physics, University of Basel, CH-4056 Basel,
     Switzerland
\item[\lsu] Louisiana State University, Baton Rouge, LA 70803, USA
\item[\beijing] Institute of High Energy Physics, IHEP, 
  100039 Beijing, China$^{\triangle}$ 
\item[\berlin] Humboldt University, D-10099 Berlin, FRG$^{\S}$
\item[\bologna] University of Bologna and INFN-Sezione di Bologna, 
     I-40126 Bologna, Italy
\item[\tata] Tata Institute of Fundamental Research, Bombay 400 005, India
\item[\ne] Northeastern University, Boston, MA 02115, USA
\item[\bucharest] Institute of Atomic Physics and University of Bucharest,
     R-76900 Bucharest, Romania
\item[\budapest] Central Research Institute for Physics of the 
     Hungarian Academy of Sciences, H-1525 Budapest 114, Hungary$^{\ddag}$
\item[\mit] Massachusetts Institute of Technology, Cambridge, MA 02139, USA
\item[\debrecen] KLTE-ATOMKI, H-4010 Debrecen, Hungary$^\P$
\item[\florence] INFN Sezione di Firenze and University of Florence, 
     I-50125 Florence, Italy
\item[\cern] European Laboratory for Particle Physics, CERN, 
     CH-1211 Geneva 23, Switzerland
\item[\wl] World Laboratory, FBLJA  Project, CH-1211 Geneva 23, Switzerland
\item[\geneva] University of Geneva, CH-1211 Geneva 4, Switzerland
\item[\hefei] Chinese University of Science and Technology, USTC,
      Hefei, Anhui 230 029, China$^{\triangle}$
\item[\seft] SEFT, Research Institute for High Energy Physics, P.O. Box 9,
      SF-00014 Helsinki, Finland
\item[\lausanne] University of Lausanne, CH-1015 Lausanne, Switzerland
\item[\lecce] INFN-Sezione di Lecce and Universit\'a Degli Studi di Lecce,
     I-73100 Lecce, Italy
\item[\lyon] Institut de Physique Nucl\'eaire de Lyon, 
     IN2P3-CNRS,Universit\'e Claude Bernard, 
     F-69622 Villeurbanne, France
\item[\madrid] Centro de Investigaciones Energ{\'e}ticas, 
     Medioambientales y Tecnolog{\'\i}cas, CIEMAT, E-28040 Madrid,
     Spain${\flat}$ 
\item[\milan] INFN-Sezione di Milano, I-20133 Milan, Italy
\item[\moscow] Institute of Theoretical and Experimental Physics, ITEP, 
     Moscow, Russia
\item[\naples] INFN-Sezione di Napoli and University of Naples, 
     I-80125 Naples, Italy
\item[\cyprus] Department of Natural Sciences, University of Cyprus,
     Nicosia, Cyprus
\item[\nymegen] University of Nijmegen and NIKHEF, 
     NL-6525 ED Nijmegen, The Netherlands
\item[\caltech] California Institute of Technology, Pasadena, CA 91125, USA
\item[\perugia] INFN-Sezione di Perugia and Universit\'a Degli 
     Studi di Perugia, I-06100 Perugia, Italy   
\item[\cmu] Carnegie Mellon University, Pittsburgh, PA 15213, USA
\item[\prince] Princeton University, Princeton, NJ 08544, USA
\item[\rome] INFN-Sezione di Roma and University of Rome, ``La Sapienza",
     I-00185 Rome, Italy
\item[\peters] Nuclear Physics Institute, St. Petersburg, Russia
\item[\salerno] University and INFN, Salerno, I-84100 Salerno, Italy
\item[\ucsd] University of California, San Diego, CA 92093, USA
\item[\santiago] Dept. de Fisica de Particulas Elementales, Univ. de Santiago,
     E-15706 Santiago de Compostela, Spain
\item[\sofia] Bulgarian Academy of Sciences, Central Lab.~of 
     Mechatronics and Instrumentation, BU-1113 Sofia, Bulgaria
\item[\korea] Center for High Energy Physics, Adv.~Inst.~of Sciences
     and Technology, 305-701 Taejon,~Republic~of~{Korea}
\item[\alabama] University of Alabama, Tuscaloosa, AL 35486, USA
\item[\utrecht] Utrecht University and NIKHEF, NL-3584 CB Utrecht, 
     The Netherlands
\item[\purdue] Purdue University, West Lafayette, IN 47907, USA
\item[\psinst] Paul Scherrer Institut, PSI, CH-5232 Villigen, Switzerland
\item[\zeuthen] DESY, D-15738 Zeuthen, 
     FRG
\item[\eth] Eidgen\"ossische Technische Hochschule, ETH Z\"urich,
     CH-8093 Z\"urich, Switzerland
\item[\hamburg] University of Hamburg, D-22761 Hamburg, FRG
\item[\taiwan] National Central University, Chung-Li, Taiwan, China
\item[\tsinghua] Department of Physics, National Tsing Hua University,
      Taiwan, China
\item[\S]  Supported by the German Bundesministerium 
        f\"ur Bildung, Wissenschaft, Forschung und Technologie
\item[\ddag] Supported by the Hungarian OTKA fund under contract
numbers T019181, F023259 and T024011.
\item[\P] Also supported by the Hungarian OTKA fund under contract
  numbers T22238 and T026178.
\item[$\flat$] Supported also by the Comisi\'on Interministerial de Ciencia y 
        Tecnolog{\'\i}a.
\item[$\sharp$] Also supported by CONICET and Universidad Nacional de La Plata,
        CC 67, 1900 La Plata, Argentina.
\item[$\diamondsuit$] Also supported by Panjab University, Chandigarh-160014, 
        India.
\item[$\triangle$] Supported by the National Natural Science
  Foundation of China.
\item[\dag] Deceased.
\end{list}
}
\vfill






\newpage
\begin{table}[htpb]
\begin{center}
\bigskip
\begin{tabular}{|c|c|c|} \hline
Category                 &      Muon channel      &   Electron channel    \\ \hline
 Observed Events          &      450               &   360                 \\ \hline
Jet misidentification    &      153$\pm$7         &   147$\pm$7           \\ 
Misidentified hadrons    &      188$\pm$8         &   101$\pm$6           \\ 
Photon conversions        &       -                &     2$\pm$1           \\ 
Dalitz decays            &       -                &    35$\pm$4           \\ 
Decays in flight         &       58$\pm$4         &        -              \\ \hline
Total Background         &      399$\pm$11        &   285$\pm$10          \\ \hline 
\end{tabular}
\label{tab:leps}
\caption[]{ Summary of the number of selected events and the backgrounds for the lepton
analysis. The quoted errors are statistical only. }
\end{center}

\end{table}

\begin{table}[htpb]
\begin{center}
\bigskip
\begin{tabular}{|c|c|c|} \hline
Category                   &  Muon channel       & Electron channel    \\ \hline
$\varepsilon^{c}$($\%$)       & 0.40 $\pm$ 0.04   & 0.54 $\pm$ 0.05   \\ \hline
$\varepsilon^{b}$($\%$)       & 0.97 $\pm$ 0.10   & 1.81 $\pm$ 0.14   \\ \hline
\end{tabular}
\label{tab:effs}
\caption[]{ Summary of selection efficiencies for $\rm{g} \ra \ccbar$ and
$\rm{g} \ra \bbbar$ events for the lepton analysis. The quoted errors are due to Monte Carlo
 statistics only. }
\end{center}
\end{table}
\begin{table}[htpb]
\begin{center}
\bigskip
\begin{tabular}{|c|c|c|} \hline
Source                  &     
 $\delta(\overline{n}^{\mu}_{\rm{g} \ra \ccbar})/\overline{n}^{\mu}_{\rm{g}
 \ra \ccbar}(\%)$      &
$\delta(\overline{n}^{\rm{e}}_{\rm{g} \ra \ccbar})/\overline{n}^{\rm{e}}_{\rm{g}
 \ra \ccbar}(\%)$ 
\\ \hline
Monte Carlo statistics                &      10.1          &  9.5            \\ 
Lepton efficiency            &      6.8          &  7.8            \\ 
Background simulation        &     10.0          &  15.0           \\
Track smearing               &      1.9          &  3.5            \\ \hline
Total experimental errors    &     15.9          & 19.7            \\ \hline
Br($\rm{b} \ra \ell$)=$0.105 \pm 0.005$            &     9.8    & 4.1      \\ 
Br($\rm{b} \ra c \ra \ell$)=$0.093 \pm 0.007$   &     8.9    &  2.1      \\  
Br($\rm{c} \ra \ell$)=$0.098 \pm 0.005$            &     13.7   & 9.6      \\ 
$R_{\rm{b}}$  = 0.2017 $\pm$ 0.0040             &      4.0   & 2.9      \\ \hline
Branching ratio errors                    &     19.5   & 11.0     \\ \hline 
$\Lambda_{LLA}$= 0.30 $\pm$ 0.03          &      4.3   &  4.9     \\ 
Symmetric parameter $b=0.76\pm$0.08      &      2.0   &  2.6     \\ 
$\sigma_{q}$= 0.39$\pm$0.03               &      2.6   &  2.6     \\
$\epsilon_{b} =-0.0056\pm0.0024$       &      1.6   &  1.6     \\ 
$\epsilon_{c}=-0.05\pm0.02$          &      1.1   &  1.1  \\ 
Semi-leptonic Decay model                  &      8.2   &  9.3  \\ \hline
Modelling errors                     &     10.0   &  11.3 \\ \hline \hline
Total                                      &     27.1 [15.8]   & 
 25.2 [15.8] \\ \hline
\end{tabular}
\label{tab:sysl}
\caption[]{Summary of systematic errors for the muon and electron
analysis. The correlated error due
 to branching ratios and QCD modelling is given in square brackets in the
 last line.}
\end{center}
\end{table}

\begin{table}[htpb]
\begin{center}
\begin{tabular}{|c|r|} \hline
Event Category           &   Efficiency(\%)      \\ \hline
Data                     & $\varepsilon_D =$\phantom{0}  1.509 $\pm$ 0.009   \\ 
$\it N$ Sample          &  $\varepsilon_N =$\phantom{0}  1.411 $\pm$ 0.006   \\ 
$\it C$ Sample          &  $\varepsilon_C =$\phantom{0}  4.401 $\pm$ 0.082   \\ 
$\it B$ Sample          &  $\varepsilon_B =$ 12.967 $\pm$ 0.470   \\ \hline
\end{tabular}
\bigskip
\caption[]{ Summary of the selection efficiencies 
in data and Monte Carlo for the event shape analysis.
 The quoted errors are due to Monte Carlo statistics only.}
\label{tab:nns}
\end{center}
\end{table}
\bigskip

\begin{table}[htpb]
\begin{center}
\begin{tabular}{|c|r|} \hline
Source                  &     
$ \delta(\ngbcc)/\ngbcc(\%)$  
\\ \hline
MC statistics                        &    9.2 \\ 
$R_{\rm{b}} =0.2017 \pm 0.0040 $             &    14.4 \\ 
Track smearing                       &    2.7 \\
Energy calibration                   &   12.5 \\
$\Lambda_{LLA}= 0.30 \pm 0.03$     &    4.8 \\ 
$\sigma_{q}=0.39\pm0.03$          &    7.6 \\
Symmetric parameter $b=0.76\pm0.08$  &    1.7 \\ 
$\epsilon_{b}=-0.0056 \pm 0.0024$  &    2.9 \\ 
$\epsilon_{c}= -0.05 \pm 0.02$     &    3.8 \\ \hline
Total                                &   23.7 \\ \hline
\end{tabular}
\bigskip
\caption[]{Summary of systematic errors for the
event shape analysis. }
\label{tab:evsh}
\end{center}
\end{table}
\begin{table}[htpb]
\begin{center}
\bigskip
\begin{tabular}{|c|c|c|} \hline
\multicolumn{3}{|c|}{ Experimental measurements of
$ \ngbcc$} \\ \hline
Expt. & Ref. & $ \ngbcc$ ($\%$)  \\ \hline
L3  & This paper & $2.45\pm0.29\pm0.53$      \\ \hline
OPAL & 8 & $3.20\pm0.21\pm0.38$ \\ \hline
ALEPH & 9 & $3.23\pm0.48\pm0.53$ \\ \hline
\multicolumn{3}{|c|}{ Theoretical predictions for
$\ngbcc$} \\ \hline
Model & Ref. & $ \ngbcc$ ($\%$)  \\ \hline
 LO pQCD & 1 & 0.607 \\ \hline
Resummed LO pQCD (A) & 3 & 1.35 \\ \hline
Resummed LO pQCD (B) & 4 & 2.01 \\ \hline
 HERWIG & 3 & 0.923 \\ \hline
 JETSET & 3 & 1.70 \\ \hline
ARIADNE & 3 & 2.18 \\ \hline  
\end{tabular}
\label{tab:reslts}
\caption[]{Summary of published experimental measurements on
$ \ngbcc$  and
 theoretical predictions from perturbative QCD calculations
 (pQCD) and Monte Carlo models (HERWIG, JETSET and ARIADNE).}
\end{center}
\end{table}
\clearpage

\clearpage

\clearpage

\clearpage

\begin{figure} 
 \begin{center}
   \includegraphics[width=15cm]{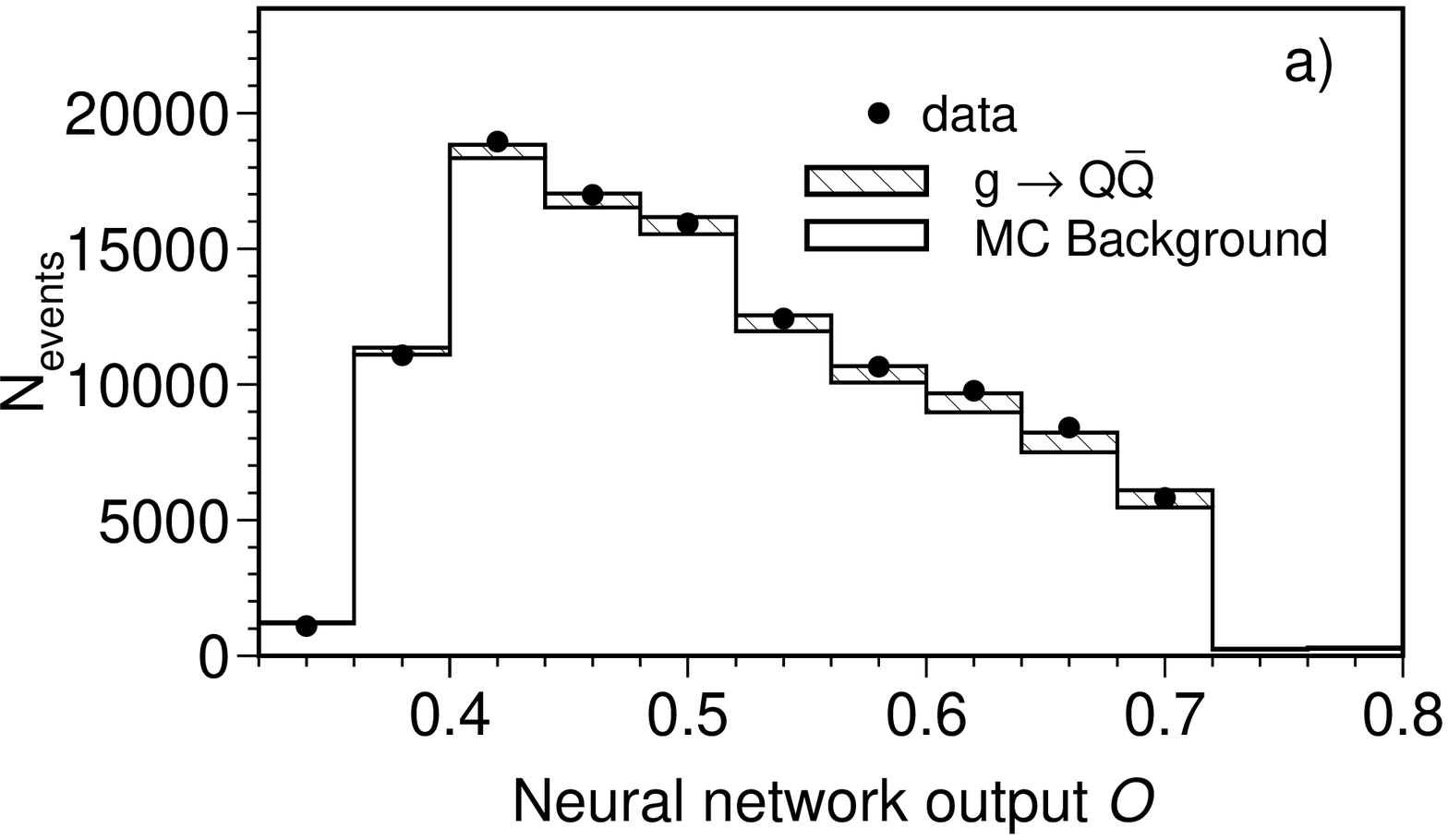}
   \includegraphics[width=15cm]{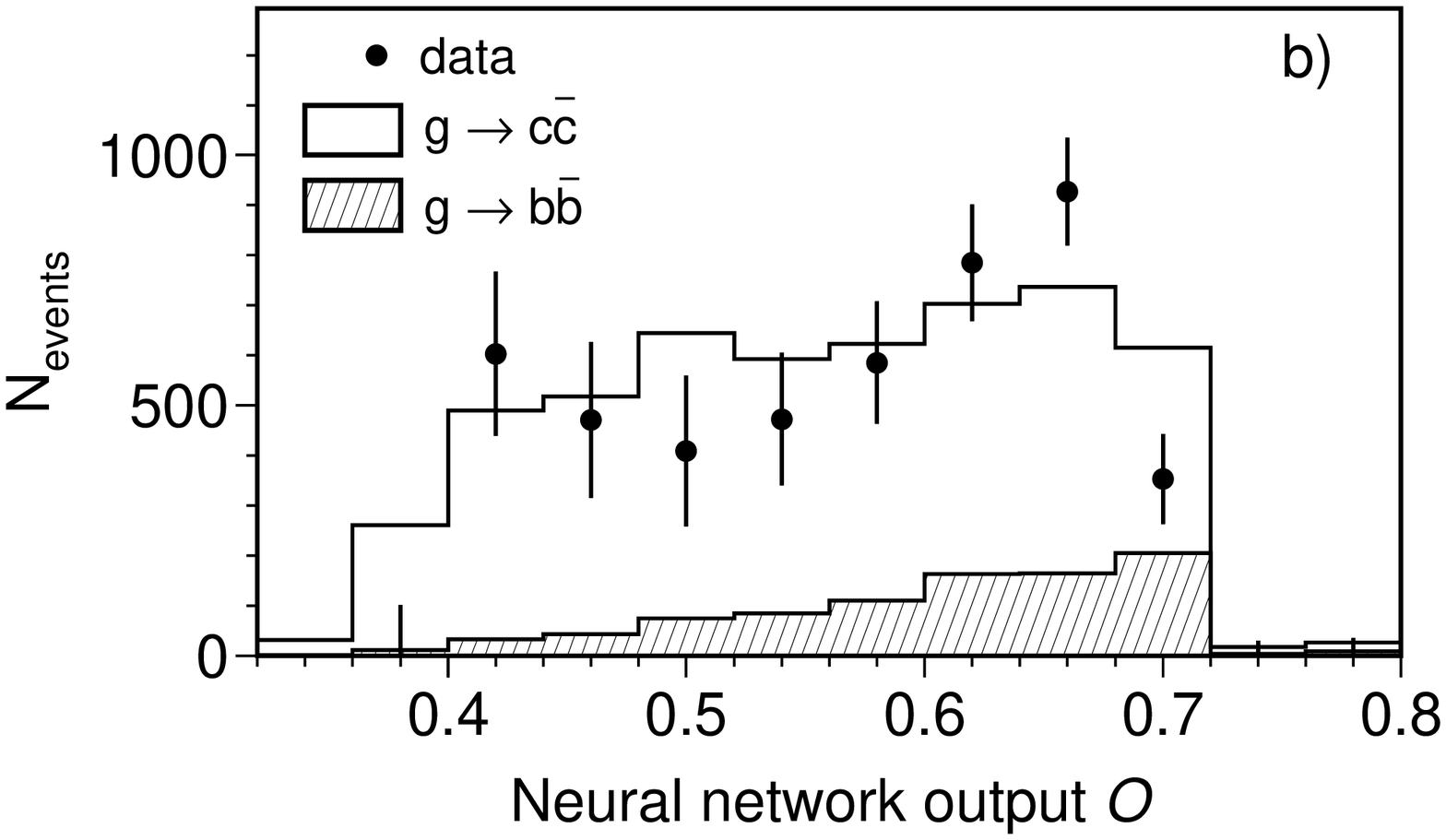}
 \end{center}

 \caption{ a) Distribution of the neural network output
for data (points), background (histogram) and 
$\rm{g} \ra \rm{Q} \overline{\rm{Q}}$ events (hatched area).
 b) background-subtracted neural network output distribution for data (points) and 
Monte Carlo $\rm{g} \ra \rm{Q}\overline{\rm{Q}}$ events (histogram). The hatched area shows the 
$\rm{g} \ra \bbbar$ contribution. 
The contributions of $\rm{g} \ra \ccbar$ Monte Carlo events in a) and b) are normalised to 
the measured value of $\ngbcc$ and the contributions 
of $\rm{g} \ra \bbbar$ events are estimated using the measured value 
 of $\ngbbb$.} 
\label{dnno}
\end{figure}


\end{document}